\documentclass[11pt]{article}
\usepackage{fullpage,graphicx,psfrag,amsmath,amsfonts,verbatim}
\usepackage[small,bf]{caption}
\usepackage{url}
\usepackage{algorithm}
\usepackage{algorithmic}
\usepackage{caption}

\usepackage[final]{pdfpages}

\begin{document}

\author{
Sam Ganzfried\\
Ganzfried Research, Cornell University\\
\texttt{sam.ganzfried@gmail.com}
}

\date{\vspace{-5ex}}

\title{Projected Exploitability Descent for Nash Equilibrium Computation in Multiplayer Imperfect-Information Games} 

\maketitle

\begin{abstract}
Many important games have more than two players and imperfect information. Existing approaches for computing Nash equilibrium, the central game-theoretic solution concept, in such games either lack scalability or obtain poor performance. In this paper we introduce a new algorithm called projected exploitability descent (PED) for approximating Nash equilibria in multiplayer games of imperfect information. The algorithm works by running projected subgradient descent minimizing a proxy for the multiplayer generalized exploitability function. The objective is nonconvex and nonsmooth, but can be represented as the sum of the maxima of linear functions, for which a subgradient can easily be computed and projected to the polytope of feasible sequence-form strategies. We explore performance of PED on a generalized version of the well-studied benchmark game three-player Kuhn poker. No prior exact algorithms scale to the version of the game with deck size larger than 4, and we compare performance to the popular algorithms of fictitious play (FP) and counterfactual regret minimization (CFR). We find that PED obtains a consistent near-monotonic improvement throughout all runs, though both FP and CFR perform significantly better in the initial iterations. This inspires a hybrid algorithm FP-PED that runs FP for an initial burn-in period before switching to PED for stable long-run refinement. We can alternatively view this as a multi-step algorithm that runs FP as a pre-processing step to obtain a strong initialization for PED.
\end{abstract}

\section{Introduction}
\label{se:intro}
Nash equilibrium is a central solution concept in game theory. A Nash equilibrium is a vector of strategies for all players such that no player can benefit from deviating. Nash equilibrium can be computed in polynomial-time in two-player zero-sum games, but is PPAD-hard in non-zero-sum and multiplayer games. This complexity result holds both for simultaneous-move normal-form games and for sequential extensive-form games of imperfect information. In this paper we are interested in imperfect-information games with $n > 2$ players. There are two general classes of approaches for this problem: algorithms for finding exact Nash equilibrium (which do not scale to large games) and scalable approaches that attempt to compute approximations of Nash equilibrium (without convergence guarantees). In this paper we are interested in solving large games not solvable by the first class; so our primary interest will be on algorithms that obtain a strong degree of approximation of Nash equilibrium despite lacking a worst-case performance guarantee.

A \emph{normal-form game} consists of a finite set of players $N = \{1,\ldots,n\}$, a finite set of pure strategies $S_i$ for each player $i$, and a real-valued utility for each player for each strategy vector (aka \emph{strategy profile}), $u_i : \times_i S_i \rightarrow \mathbb{R}$. Let $\Sigma_i$ denote the set of mixed strategies of player $i$ (probability distributions over elements of $S_i$). If players follow mixed strategy profile $\sigma = (\sigma^{(1)},\ldots,\sigma^{(n)})$, where $\sigma^{(i)} \in \Sigma_i$, the expected payoff to player $i$ is
\small
$$
u_i(\sigma^{(1)},\ldots,\sigma^{(n)})
=
\sum_{s_1 \in S_1,\ldots,s_n \in S_n}
\sigma^{(1)}_{s_1} \cdots \sigma^{(n)}_{s_n}
u_i(s_1,\ldots,s_n).
$$ 
\normalsize
We write $u_i(\sigma) = u_i(\sigma^{(i)},\sigma^{(-i)})$, where $\sigma^{(-i)}$ denotes the vector of strategies of all players except $i$. A mixed strategy profile $\sigma^*$ is a \emph{Nash equilibrium} if for each player $i \in N$ and for all $\sigma^{(i)} \in \Sigma_i$,
$
u_i(\sigma^{*(i)},\sigma^{*(-i)})
\geq
u_i(\sigma^{(i)},\sigma^{*(-i)}).
$
Every finite game contains at least one Nash equilibrium.

Given a strategy profile $\sigma$, for player $i \in N$, define
\small
$$
\epsilon_i(\sigma)
=
\max_{s_i \in S_i}
\left[
u_i(s_i,\sigma^{(-i)})
-
u_i(\sigma^{(i)},\sigma^{(-i)})
\right] \mbox{ and}
$$
$$
\epsilon(\sigma)
=
\max_{i \in N} \epsilon_i(\sigma)
=
\max_{i \in N}
\max_{s_i \in S_i}
\left[
u_i(s_i,\sigma^{(-i)})
-
u_i(\sigma^{(i)},\sigma^{(-i)})
\right].
$$ 
\normalsize
We refer to $\epsilon(\sigma)$ as the \emph{exploitability}\footnote{Note that \emph{exploitability} is traditionally defined just for two-player zero-sum games. A two-player game is \emph{zero sum} if for all strategy profiles $(s_1,s_2)$, $u_1(s_1,s_2) + u_2(s_1,s_2) = 0.$ In two-player zero-sum games the minimax theorem states that there exists a value $v^* \in \mathbb{R}$ such that $v^* = \max_{s_i \in S_1} \min_{s_2 \in S_2} u_1(s_1,s_2) = \min_{s_2 \in S_2} \max_{s_1 \in S_1} u_1(s_1,s_2).$ The quantity $v_1 = v^*$ is called the \emph{value} of the game to player 1 ($v_2 = -v^*$ is the value of the game to player 2). A strategy profile $\sigma^* = (\sigma^{*(1)},\sigma^{*(2)})$ is a Nash equilibrium iff for each player $i$, $\sigma^{*(i)}$ obtains worst-case expected value of $v_i.$ For an arbitrary strategy $\sigma^{(1)},$ the \emph{exploitability} of $\sigma^{(1)}$ is defined as $v_1 - \min_{\sigma^{(2)} \in \Sigma_2} u_1(\sigma^{(1)},\sigma^{(2)}).$ Thus, $\sigma^{(1)}$ is a Nash equilibrium strategy for player 1 iff its exploitability is 0. We can then define the \emph{exploitability of strategy profile} $(\sigma^{(1)},\sigma^{(2)})$ as the maximum of the exploitability of $\sigma^{(1)}$ and the exploitability of $\sigma^{(2)}$. We can further extend this to obtain our definition of the \emph{exploitability of a strategy profile in an $n$-player normal-form game}.} of strategy profile $\sigma$, which is the standard metric for evaluating the degree to which $\sigma$ approximates a Nash equilibrium. Note that $\epsilon(\sigma) = 0$ iff $\sigma$ is an exact Nash equilibrium, and that $\epsilon(\sigma)$ is a nonconvex and nondifferentiable function of $\sigma.$ 

Although exploitability is the standard metric for evaluating approximate Nash equilibria, its outer maximization makes it difficult to optimize directly using gradient-based methods.
Recent work proposed a gradient-based approach for approximating Nash equilibria in multiplayer normal-form games~\cite{Wang25:Approximating}. Their approach is based on a distance-to-equilibrium objective called \emph{NashD}, which measures the aggregate unilateral incentive for players to deviate (in contrast to $\epsilon$ which uses the maximum). For a mixed strategy profile $\sigma = (\sigma^{(1)},\ldots,\sigma^{(n)})$, they define
\[
\mathrm{NashD}(\sigma)
=
\sum_{i \in N}
\left[
\max_{s_i \in S_i}
u_i(s_i,\sigma^{(-i)})
-
u_i(\sigma^{(i)},\sigma^{(-i)})
\right],
\]
which equals $\sum_{i \in N} \epsilon_i(\sigma).$
Like exploitability, $\mathrm{NashD}(\sigma)\ge 0$, with equality if and only if $\sigma$ is a Nash equilibrium.

The optimization procedure performs gradient descent directly on a parameterized representation of the strategy profile. Specifically, each player maintains an unconstrained vector $\sigma'^{(i)}$ of real-valued parameters, which is mapped to a valid mixed strategy $\sigma^{(i)}$ using the softmax transformation
\[
\sigma^{(i)}(s_{ik})
=
\frac{\exp(\sigma'^{(i)}(s_{ik}))}
{\sum_{j=1}^{|S_i|}
\exp(\sigma'^{(i)}(s_{ij}))}.
\]
The parameters are then updated iteratively according to
$
\sigma'^{(i)}
\leftarrow
\sigma'^{(i)}
-
\alpha
\nabla_{\sigma'^{(i)}}
\mathrm{NashD}(\sigma).
$

Because the NashD objective contains a maximization over unilateral deviations, it is generally nondifferentiable. In particular, if multiple pure strategies simultaneously achieve the maximal payoff against the current opponents' strategy profile, the objective is not differentiable at that point. Consequently, the update is more accurately interpreted as a projected subgradient step rather than ordinary gradient descent.

The subgradient is computed by first identifying, for each player $i$, a pure strategy
\[
s_i^\ast \in \arg\max_{s_i \in S_i} u_i(s_i,\sigma^{-i})
\]
that attains the maximum in the corresponding NashD term. Locally, the objective is then treated as the active branch
\[
u_i(s_i^\ast,\sigma^{-i}) - u_i(\sigma^i,\sigma^{-i}),
\]
which is one of the linear pieces defining the maximum. Since expected utilities are multilinear functions of the mixed strategies, the active piece has an explicit gradient with respect to the mixed-strategy variables. Applying the chain rule through the softmax parameterization then yields the gradient with respect to the unconstrained optimization variables.

If multiple pure strategies simultaneously maximize the payoff, the objective is nondifferentiable, but the gradient of any active branch is a valid subgradient of the maximum function. Thus, each iteration consists of selecting a current best response for every player, computing the corresponding subgradient of NashD, and taking a descent step in the unconstrained parameter space. Experimental results demonstrated that the method significantly outperformed fictitious play and regret matching on random games and games from the GAMUT benchmark suite~\cite{Nudelman04:Run}.

\section{Imperfect-Information Games}
\label{se:imp-info}
Imperfect-information games are modeled using extensive-form game trees, where play proceeds from the root node to a terminal leaf node at which point all players receive payoffs. Each nonterminal node has an associated player (possibly chance) that makes the decision at that node. These nodes are partitioned into information sets, where the player whose turn it is to move cannot distinguish among the states in the same information set. Therefore, within a given information set, a player must choose actions according to the same probability distribution at each state contained in that information set. If no player forgets information that they previously knew, the game is said to have perfect recall. A behavioral strategy for player $i$ specifies a probability distribution over actions at each information set belonging to player $i$. 

Rather than operate on the full normal-form strategy space, which is exponential in the size of the game tree, we use the sequence-form representation~\cite{Koller94:Fast}. Under perfect recall, behavioral strategies admit an equivalent sequence-form representation based on realization plans. Rather than optimize directly over behavioral strategies, we work with their equivalent sequence-form representation. The sequence-form representation is polynomial in the size of the game tree and defines a convex feasible region described by linear constraints, making it well suited for projected first-order optimization methods.

A realization plan $x_i$ assigns to each action sequence of player $i$ the probability that the sequence is played under player $i$'s behavioral strategy. For each player $i$, let $d_i$ denote the number of action sequences. Define the sequence-form polytope 
$X_i = \{x_i \in \mathbb{R}^{d_i} : E_i x_i = e_i,\; x_i \ge 0\},$
where $E_i$ is the sequence-form constraint matrix and $e_i$ is the corresponding flow vector. The linear constraints enforce consistency of realization
probabilities across the game tree, while the nonnegativity constraints
ensure that all realization probabilities are valid. Thus, $X_i$ is a
convex polytope representing all feasible realization plans of player
$i$. For an $n$-player game, let $x = (x_1,\ldots,x_n)$
denote a joint realization-plan profile. Expected utilities are multilinear functions of the realization plans,
which we denote by
$
u_i(x_1,\ldots,x_n).
$

This multilinear structure is the key property that enables efficient
subgradient computation in the projected exploitability descent
algorithm described in the next section.

\section{Projected Exploitability Descent for Sequence-Form Games}
\label{se:algorithm}
Motivated by the subgradient-based approach for approximating Nash
equilibria in multiplayer normal-form games~\cite{Wang25:Approximating},
we develop a method called projected exploitability descent (PED) for
multiplayer imperfect-information games in sequence form. The central idea is to optimize directly over the sequence-form realization-plan 
polytope rather than an unconstrained softmax parameterization of the normal-form strategy space.
This allows the algorithm to exploit the compact sequence-form representation while maintaining feasibility by projecting each iterate onto the realization-plan polytope. 
We first define exploitability in sequence form, and then introduce the smoother
aggregate deviation objective that PED optimizes.

For a realization-plan profile $x$, define exploitability by

\[
\epsilon(x)
=
\max_{i=1,\ldots,n}
\max_{z_i\in X_i}
\left[
u_i(z_i,x_{-i})
-
u_i(x_i,x_{-i})
\right].
\]

As in the normal-form setting, $\epsilon(x)\ge 0$, with
equality iff $x$ is a Nash equilibrium. Although
exploitability is the standard metric for evaluating the
quality of an approximate Nash equilibrium, its outer
maximization makes it difficult to optimize directly using
projected subgradient methods.

For a realization-plan profile $x$, define the unilateral
deviation incentive of player $i$ by
\[
\epsilon_i(x)
=
\max_{z_i \in X_i}
[u_i(z_i,x_{-i})
-
u_i(x_i,x_{-i})],
\]
and
$\Phi(x) = \sum_{i=1}^n \epsilon_i(x).$ The function $\Phi(x)$ measures the aggregate unilateral
incentive for all players to deviate from the current
realization-plan profile. Like exploitability, $\Phi(x)\ge 0$,
with equality iff $x$ is a Nash equilibrium.
However, $\Phi$ replaces the outer maximization over
players with a summation, making it a substantially smoother surrogate objective
while preserving the same global minimizers, since
both functions are nonnegative and equal zero if and only if
$x$ is a Nash equilibrium. Consequently, $\Phi$ is
substantially better suited to projected subgradient
optimization than exploitability itself. Throughout the
remainder of this section, we derive our algorithm by
minimizing $\Phi$, while exploitability remains the primary
metric for evaluating solution quality.

Evaluating both exploitability and $\Phi$ requires computing
a best response for each player against the current
realization-plan profile. In sequence form, this can be
accomplished by solving a linear program. For fixed opponent
profile $x_{-i}$, player $i$'s best response is obtained by
solving
$b_i := b_i(x_{-i})
\in
\arg\max_{z_i \in X_i}
u_i(z_i,x_{-i}).
$
The objective is linear in $z_i$ because the opponents'
realization plans are fixed, while the feasible region $X_i$
is defined by linear equality and inequality constraints.
Thus, each best-response computation reduces to a linear
program, which can be solved efficiently using standard
optimization techniques.
Thus,
$
\epsilon_i(x)
=
u_i(b_i,x_{-i})
-
u_i(x_i,x_{-i}).
$
Although $\Phi$ is nondifferentiable because each $\epsilon_i(x)$ contains a maximization over feasible realization plans, the objective is locally represented by a differentiable function once the current best responses are fixed. Specifically, for fixed best responses $b_1,\ldots,b_n$, each maximization is replaced by its active linear piece. This is directly analogous to the subgradient construction used for the normal-form NashD objective, but with realization plans replacing mixed strategies.

Because each $\epsilon_i(x)$ is defined as the maximum of linear functions in $x_i$, fixing current best responses $b_1,\ldots,b_n$ yields a valid subgradient of $\Phi$ with respect to player $i$'s realization plan:

\[
g_i
\in
\partial_{x_i}\Phi(x)
=
\sum_{j=1}^n
\nabla_{x_i}
\left[
u_j(b_j,x_{-j})
-
u_j(x_j,x_{-j})
\right].
\]
The first term represents the change in each player's
best-response payoff as player $i$ changes its realization
plan, while the second term represents the corresponding
change in the payoff obtained by the current realization-plan
profile. Because expected utilities are multilinear in the
realization plans, each gradient is obtained by
differentiating the corresponding multilinear payoff
expression with respect to $x_i$.

Thus, after computing a best response for each player, PED
forms a subgradient of the aggregate deviation objective
$\Phi$ at the current realization-plan profile and takes a
descent step in the negative subgradient direction. Because
this step generally leaves the sequence-form feasible region,
it is followed by a Euclidean projection back onto the
polytope $X_i$.
The projected update is then
$
x_i^{t+1}
=
\Pi_{X_i}
\left(
x_i^t
-
\alpha_t g_i^t
\right),
$
where 
$
\Pi_{X_i}(y)
=
\arg\min_{x_i \in X_i}
\|x_i-y\|_2^2.
$
The projection is necessary because a subgradient step need
not satisfy the sequence-form constraints. Projecting onto
$X_i$ restores feasibility by finding the closest valid
realization plan in Euclidean distance. Consequently, every
iterate produced by PED remains a feasible realization-plan
profile.

Algorithm~\ref{alg:ped} summarizes the complete projected
exploitability descent procedure. Each iteration first
computes a best response for every player with respect to the
current realization-plan profile, then constructs the
corresponding subgradients of $\Phi$, performs a projected
subgradient step, and repeats until the prescribed iteration
limit is reached.

\begin{algorithm}[t]
\caption{Projected Exploitability Descent (PED)}
\label{alg:ped}
\begin{algorithmic}
\STATE \textbf{Input:} Initial realization-plan profile
$x^0\in X_1\times\cdots\times X_n$,
learning-rate schedule $\{\alpha_t\}_{t=0}^{T-1}$.
\STATE $x \gets x^0$
\FOR{$t=0,\ldots,T-1$}
    \FOR{each player $i=1,\ldots,n$}
        \STATE Compute a sequence-form best response
        \[
        b_i\in
        \arg\max_{z_i\in X_i}
        u_i(z_i,x_{-i}^t).
        \]
    \ENDFOR
    \FOR{each player $i=1,\ldots,n$}
        \STATE Compute a subgradient of $\Phi$
        \[
        g_i\in
        \partial_{x_i}\Phi(x^t).
        \]
        \STATE Take a projected subgradient step
        \[
        x_i^{t+1}
        =
        \Pi_{X_i}
        \left(
        x_i^t-\alpha_t g_i
        \right).
        \]
    \ENDFOR
\ENDFOR
\STATE \textbf{Output:} Approximate Nash equilibrium realization-plan profile $x^T$.
\end{algorithmic}
\end{algorithm}

Compared to the normal-form NashD algorithm, PED introduces three principal modifications. First, optimization is performed over sequence-form realization plans rather than softmax-parameterized mixed strategies. Second, best responses are computed by solving sequence-form linear programs instead of enumerating pure strategies. Third, feasibility is maintained through Euclidean projection onto the sequence-form polytope after every subgradient step. These modifications preserve the overall first-order optimization framework of Wang et al.~\cite{Wang25:Approximating} while extending it to the compact sequence-form representation required for large imperfect-information games.

\section{Experiments}
Our primary empirical evaluation is on a generalized version of three-player Kuhn poker, a well-studied imperfect-information game that has been used as a testbed in the AAAI Computer Poker Competition (e.g.,\cite{Abou10:Using,Szafron13:Parametrized}). Three-player Kuhn poker is a simplified version of poker with a four-card deck and single round of betting; it can be generalized naturally by increasing the deck size.\footnote{Each player is dealt a single card from the deck and non-dealt cards remain unused. So in the 4-card version there is one unused card, and in the general $d$-card version there are $d-3$ unused cards.} Prior work has shown that an exact approach based on directly solving a nonlinear complementarity program formulation was initially unable to solve the 4-card version~\cite{Ganzfried26:Quadratic}; however, a subsequent improved implementation was able to solve the game in 1.1 seconds~\cite{Ganzfried26:Variable}, though it was unable to solve the 5-card version in 8 hours. Of the algorithms from version 16.4 of the Gambit software suite~\cite{Savani25b:Gambit} only the logit path-following method~\cite{Turocy10:Computing} was able to solve the 4-card version (in 2.5 minutes). Thus, we are not aware of any exact algorithm that can solve the 5-card version. This makes it a good benchmark for evaluating algorithms for approximate Nash equilibrium computation. Note that the $d$-card version of three-player Kuhn poker has $4d$ information sets and $8d$ action sequences per player.

The primary benchmark algorithms we evaluate against are fictitious play (FP)~\cite{Brown51:Iterative} and (vanilla) counterfactual regret minimization (CFR)~\cite{Zinkevich07:Regret}. These are well-studied algorithms which are guaranteed to converge to Nash equilibrium under self play in two-player zero-sum games~\cite{Robinson51:Iterative}. In games with more than two players neither algorithm has a performance guarantee with respect to Nash equilibrium convergence, though CFR is guaranteed to converge to a different solution concept called coarse-correlated equilibrium~\cite{Blum07:Learning}. Prior experiments on multiplayer normal-form games with randomly generated payoffs and games from the GAMUT benchmark suite~\cite{Nudelman04:Run} show that both algorithms perform poorly for Nash equilibrium approximation, with FP outperforming regret matching~\cite{Ganzfried25:Empirical}.\footnote{Note that CFR can be viewed as the extension of regret matching to extensive-form games of imperfect information.} Additional experiments in normal-form games further validate these results, and also demonstrate that the NashD projected subgradient descent approach significantly outperforms FP and regret matching (RM)~\cite{Wang25:Approximating}. In particular, they observe:

\begin{quote}
It is worthy to notice that FP performs fairly
well in most game instances which may explain why it was used such widely in adversarial
scenarios. On the contrast, RM cannot be regarded as a proper Nash equilibrium solver,
since most strategies it obtains are not close enough to any Nash equilibrium in general
games~\cite{Wang25:Approximating}.
\end{quote}
 
In our experiments we report the results using the average strategy at a given iteration as opposed to the final iterate strategy for both fictitious play and counterfactual regret minimization; these are the standard implementations of both methods and are consistent with the convergence guarantees for two-player zero-sum games. For PED we use a standard exponential decay learning-rate schedule: 
$$\alpha_t = 0.05 \cdot 0.95^{\left\lfloor t/200 \right\rfloor}$$ 
All methods were initialized from uniform feasible strategies. In all experiments we run all algorithms for 20,000 iterations. 

\begin{figure*}[!ht]
\centering
\begin{minipage}{0.48\textwidth}
\centering
\includegraphics[width=\textwidth]{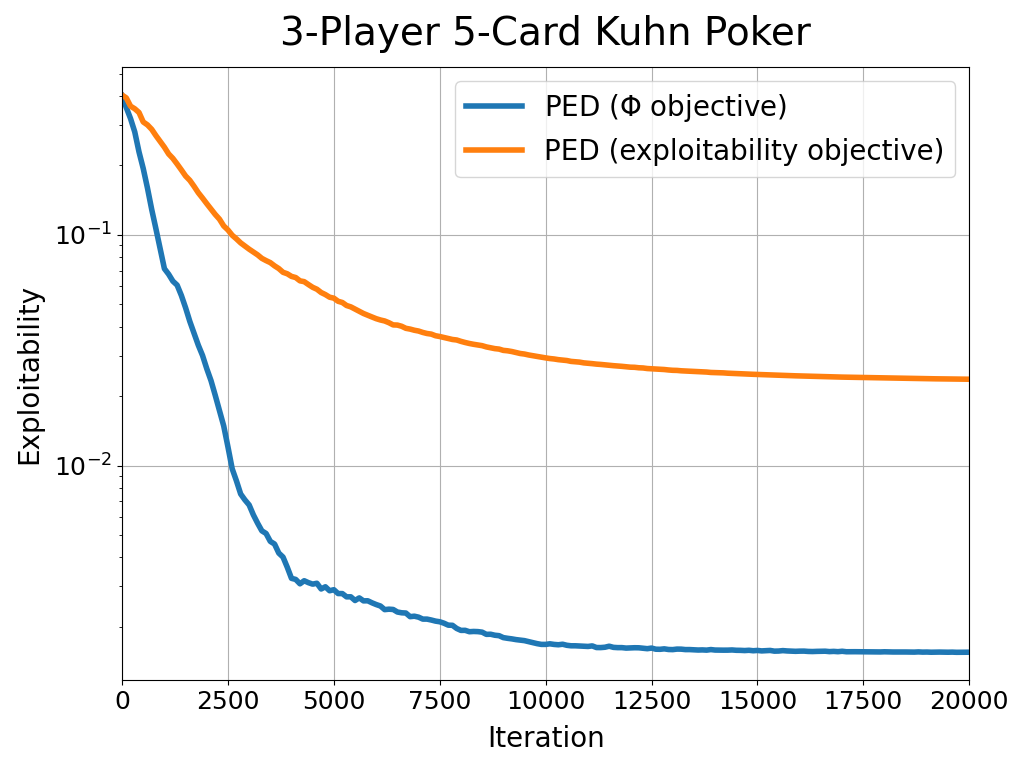}
\caption{Exploitability versus iteration.}
\label{fig:5c-expl-obj}
\end{minipage}
\hfill
\begin{minipage}{0.48\textwidth}
\centering
\includegraphics[width=\textwidth]{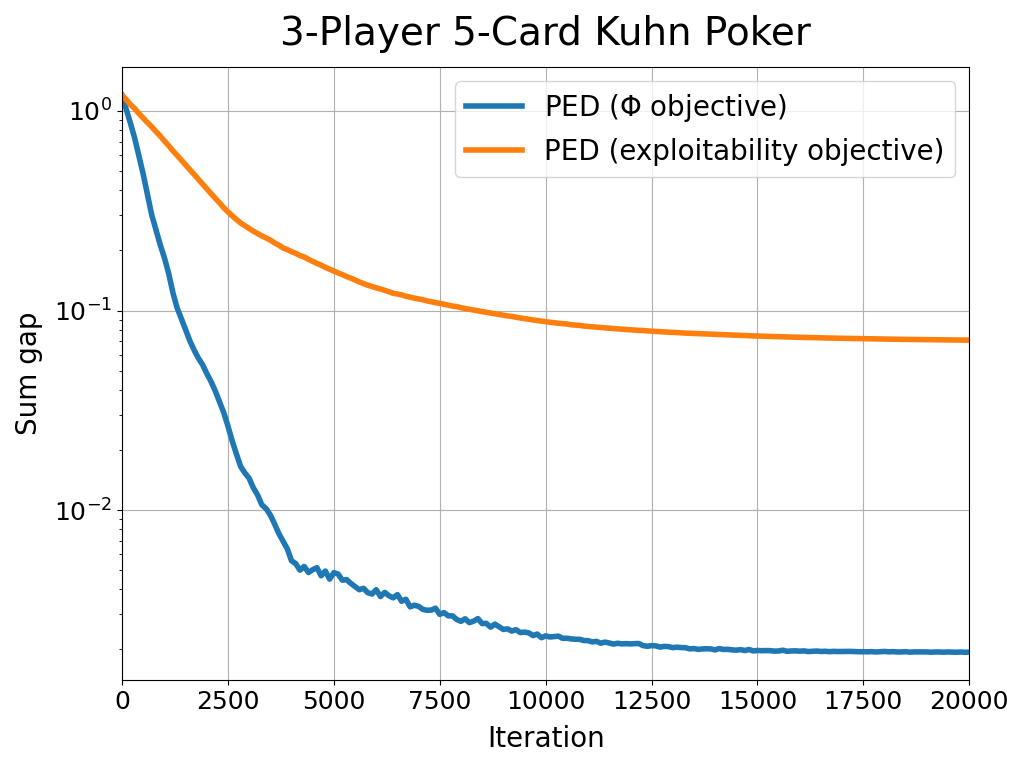}
\caption{Sum gap versus iteration.}
\label{fig:5c-sg-obj}
\end{minipage}
\end{figure*}

Before presenting our main results, we first provide further justification for the decision to minimize $\Phi$ as opposed to exploitability $\epsilon$ in our algorithm.\footnote{Note that some prior work specifically tries to minimize the $\Phi$ function as opposed to exploitability, e.g.,~\cite{Lanctot09:Monte}.} As described earlier, exploitability has nested max functions, making it significantly less smooth than $\Phi$, which replaces the outer maximization over players with a summation. While subgradient descent is designed for nonsmooth functions, its performance is better on smoother functions for which local gradient information is more useful for predicting the direction of improvement. Figure~\ref{fig:5c-expl-obj} compares exploitability when optimizing $\Phi$ versus exploitability directly in 3-player 5-card Kuhn poker. Optimizing $\Phi$ achieves a final exploitability of 0.0016 vs. 0.0236 when optimizing exploitability directly. Thus, optimizing $\Phi$ yields a 14.75-fold reduction in exploitability despite optimizing only a surrogate objective rather than exploitability directly. Figure~\ref{fig:5c-sg-obj} shows analogous results for the objective $\Phi$, which we refer to as the \emph{sum gap}. These results provide empirical support for using $\Phi$ as a smoother surrogate objective for exploitability in PED. In the remaining experiments, PED denotes our main algorithm that optimizes $\Phi$; however, we retain the terminology ``exploitability descent'' because exploitability remains the primary evaluation metric.

\begin{figure*}[!ht]
\centering
\begin{minipage}{0.48\textwidth}
\centering
\includegraphics[width=\textwidth]{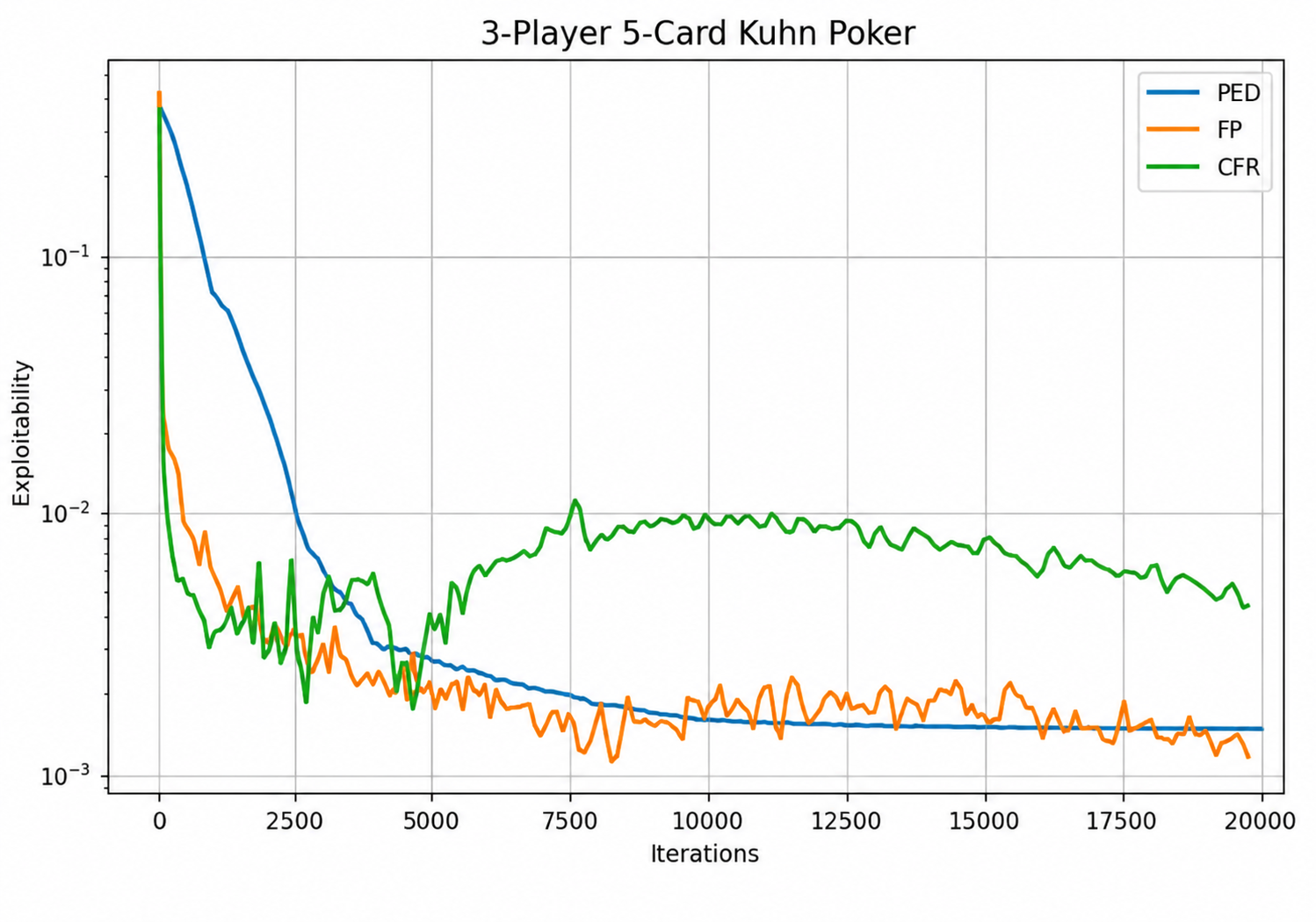}
\caption{Exploitability versus iteration.}
\label{fig:5c-expl-iter}
\end{minipage}
\hfill
\begin{minipage}{0.48\textwidth}
\centering
\includegraphics[width=\textwidth]{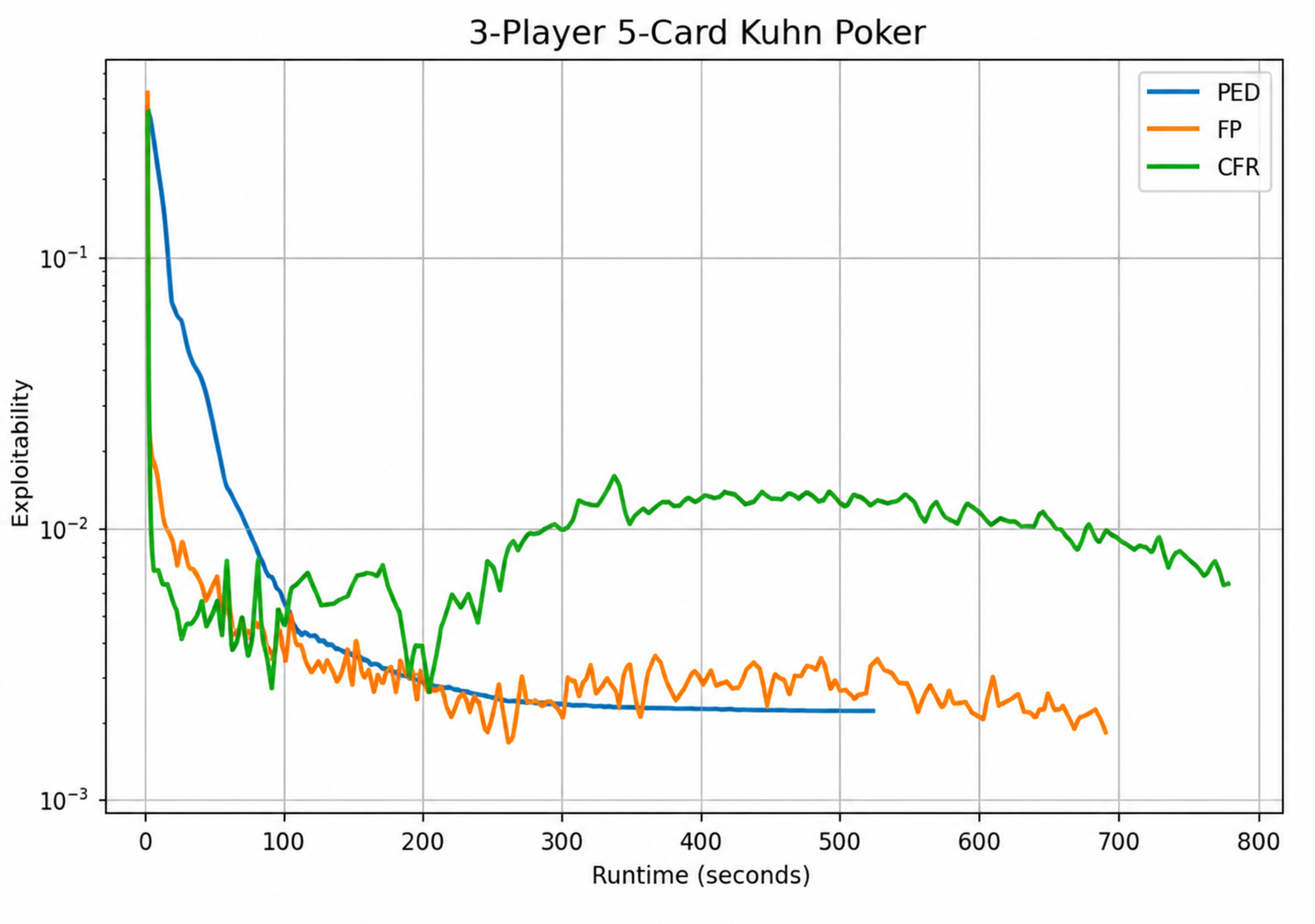}
\caption{Exploitability versus runtime.}
\label{fig:5c-expl-runtime}
\end{minipage}
\end{figure*}

\begin{figure*}[!ht]
\centering
\begin{minipage}{0.48\textwidth}
\centering
\includegraphics[width=\textwidth]{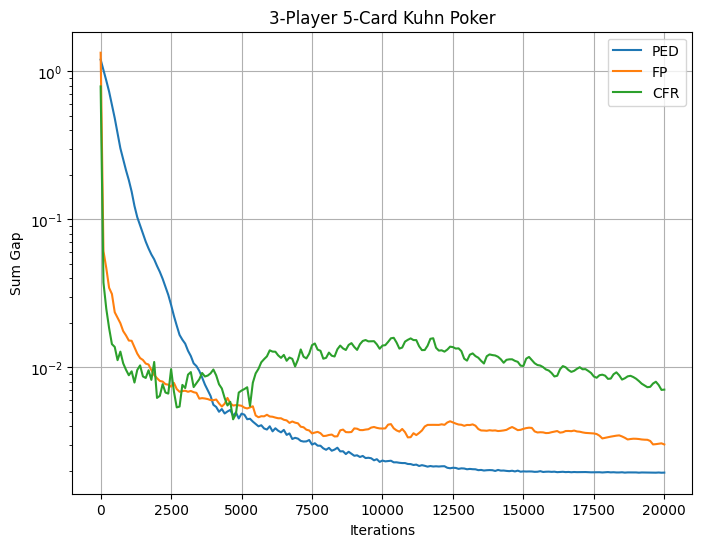}
\caption{Sum gap versus iteration.}
\label{fig:5c-sg-iter}
\end{minipage}
\hfill
\begin{minipage}{0.48\textwidth}
\centering
\includegraphics[width=\textwidth]{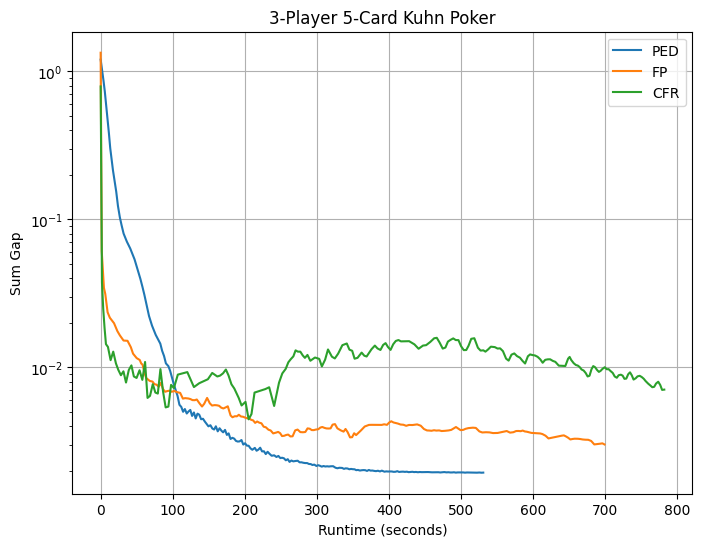}
\caption{Sum gap versus runtime.}
\label{fig:5c-sg-runtime}
\end{minipage}
\end{figure*}

\begin{figure*}[!ht]
\centering
\begin{minipage}{0.48\textwidth}
\centering
\includegraphics[width=\textwidth]{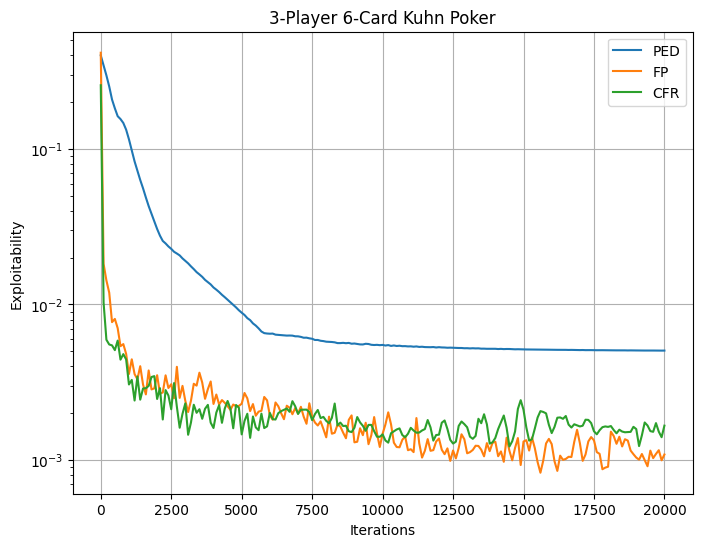}
\caption{Exploitability versus iteration.}
\label{fig:6c-expl-iter}
\end{minipage}
\hfill
\begin{minipage}{0.48\textwidth}
\centering
\includegraphics[width=\textwidth]{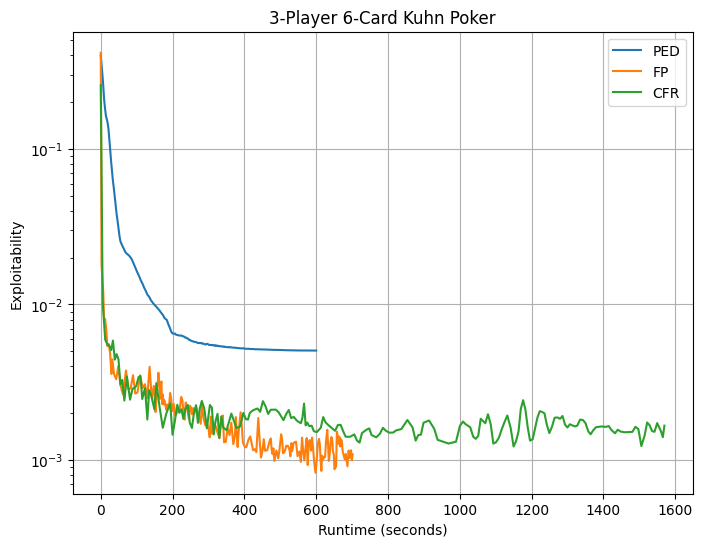}
\caption{Exploitability versus runtime.}
\label{fig:6c-expl-runtime}
\end{minipage}
\end{figure*}

\begin{figure*}[!ht]
\centering
\begin{minipage}{0.48\textwidth}
\centering
\includegraphics[width=\textwidth]{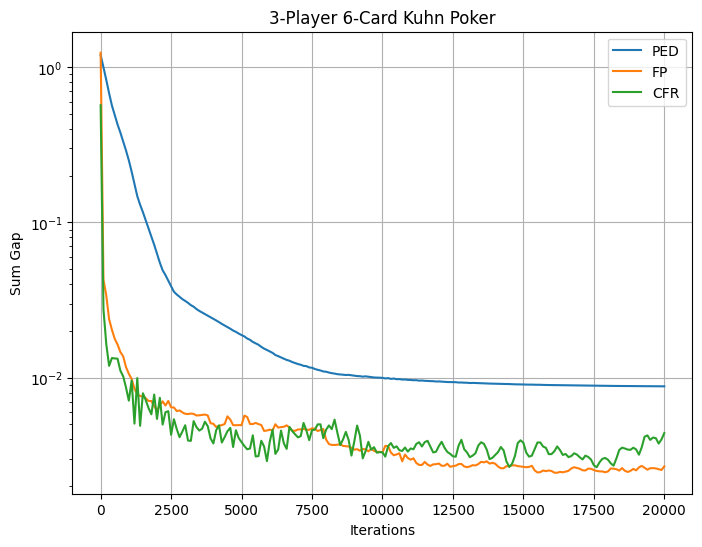}
\caption{Sum gap versus iteration.}
\label{fig:6c-sg-iter}
\end{minipage}
\hfill
\begin{minipage}{0.48\textwidth}
\centering
\includegraphics[width=\textwidth]{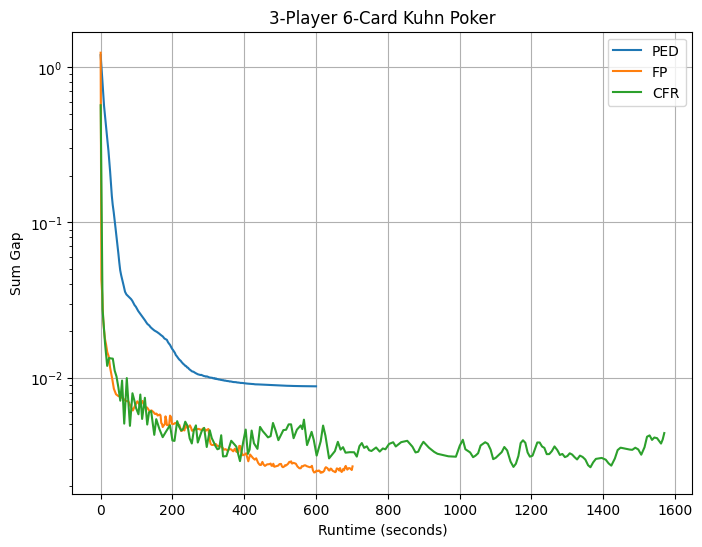}
\caption{Sum gap versus runtime.}
\label{fig:6c-sg-runtime}
\end{minipage}
\end{figure*}

\begin{figure*}[!ht]
\centering
\begin{minipage}{0.48\textwidth}
\centering
\includegraphics[width=\textwidth]{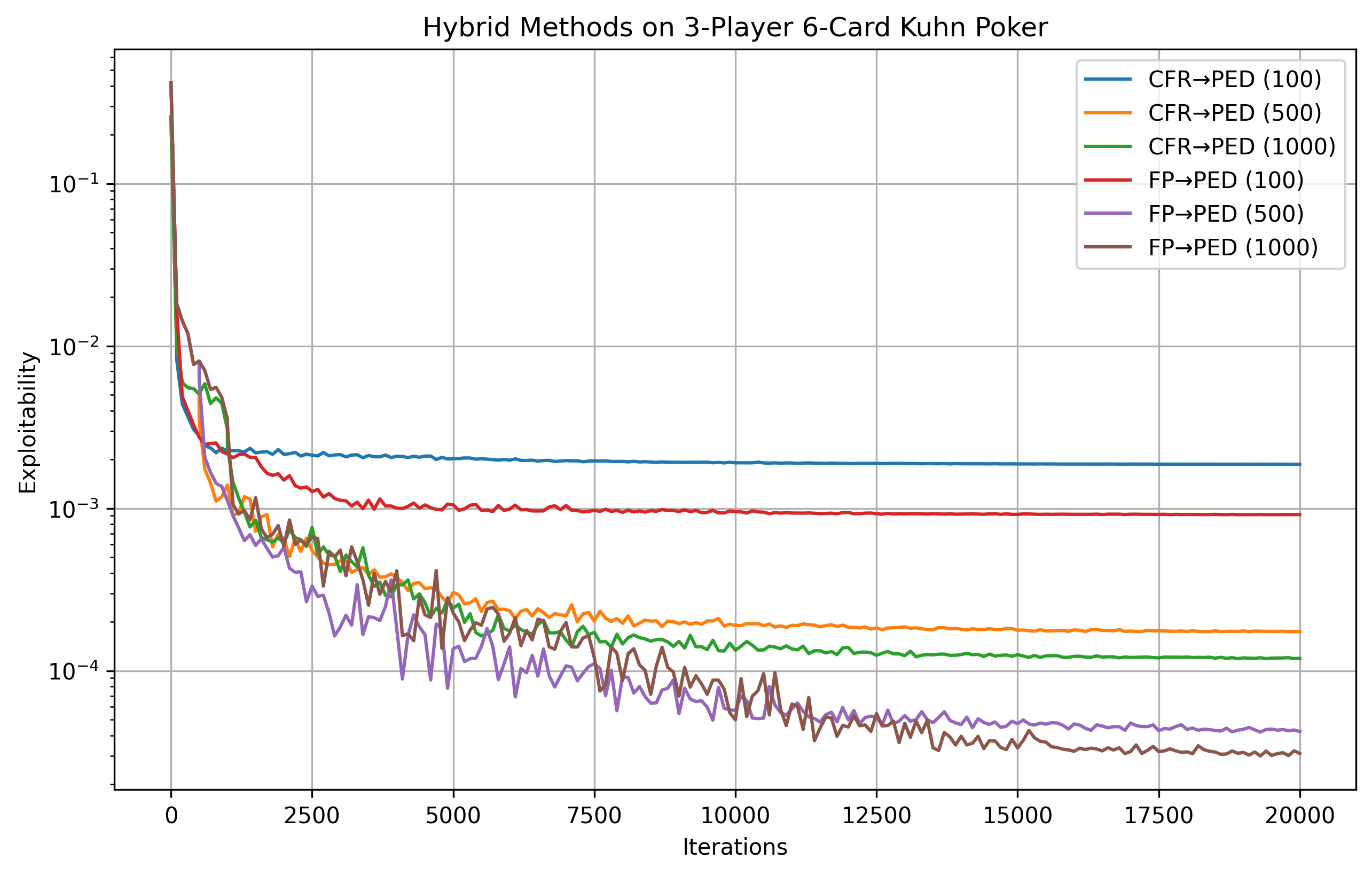}
\caption{Exploitability for hybrid methods.}
\label{fig:6c-hybrid-expl}
\end{minipage}
\hfill
\begin{minipage}{0.48\textwidth}
\centering
\includegraphics[width=\textwidth]{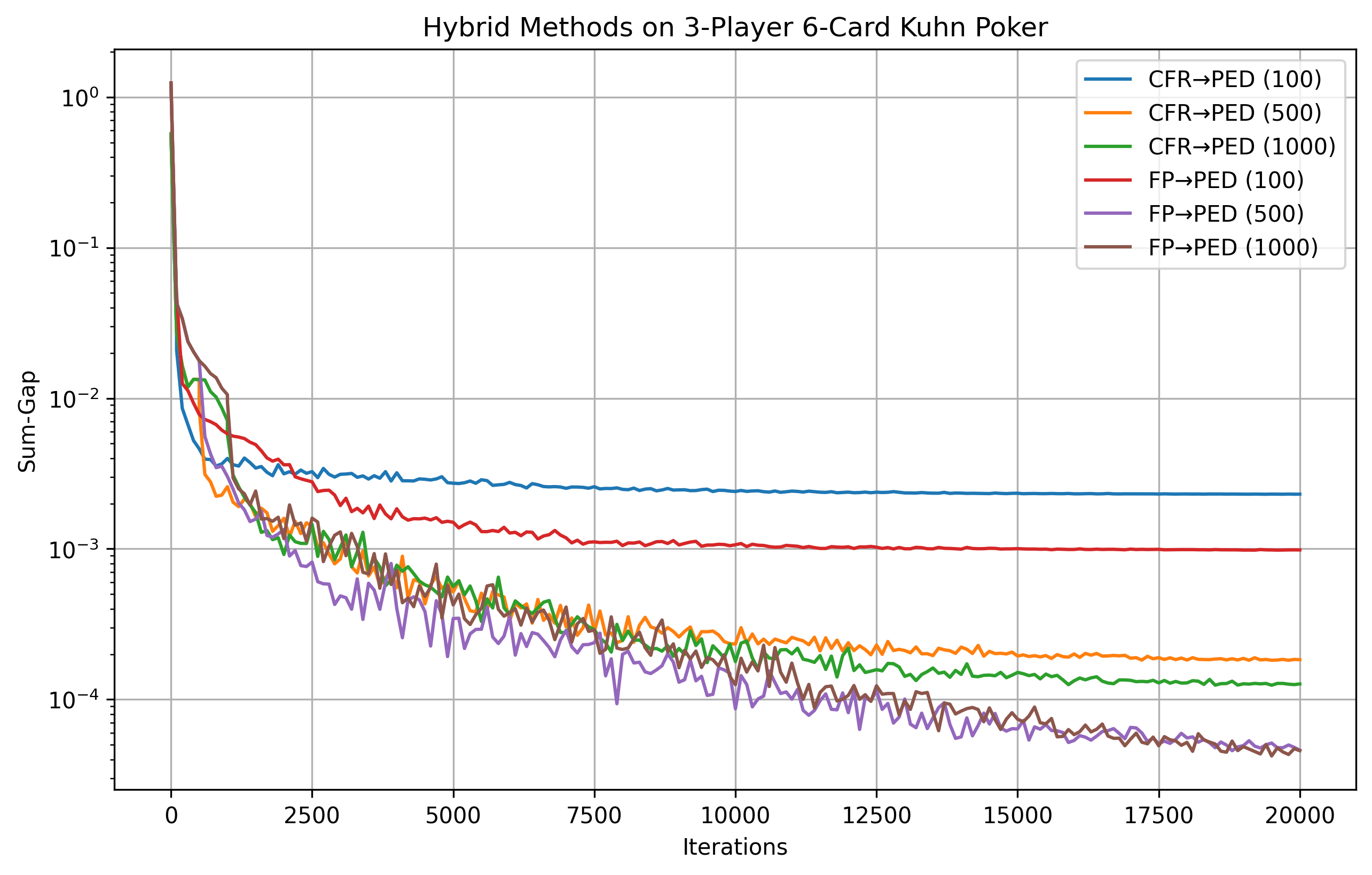}
\caption{Sum gap for hybrid methods.}
\label{fig:6c-hybrid-sg}
\end{minipage}
\end{figure*}

\begin{figure*}[!ht]
\centering
\begin{minipage}{0.48\textwidth}
\centering
\includegraphics[width=\textwidth]{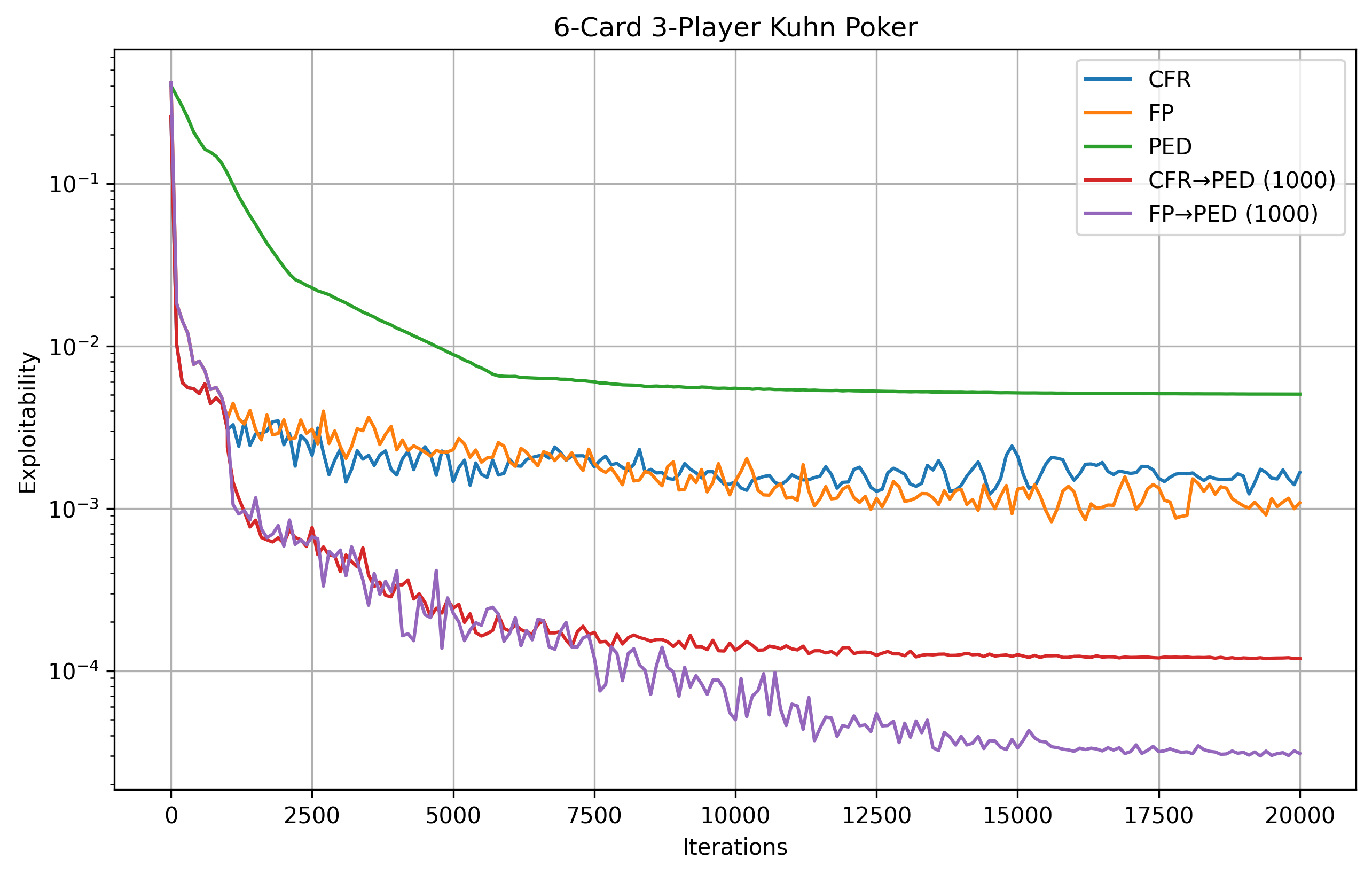}
\caption{\scriptsize Exploitability comparison with best hybrids.}
\label{fig:6c-final-expl}
\end{minipage}
\hfill
\begin{minipage}{0.48\textwidth}
\centering
\includegraphics[width=\textwidth]{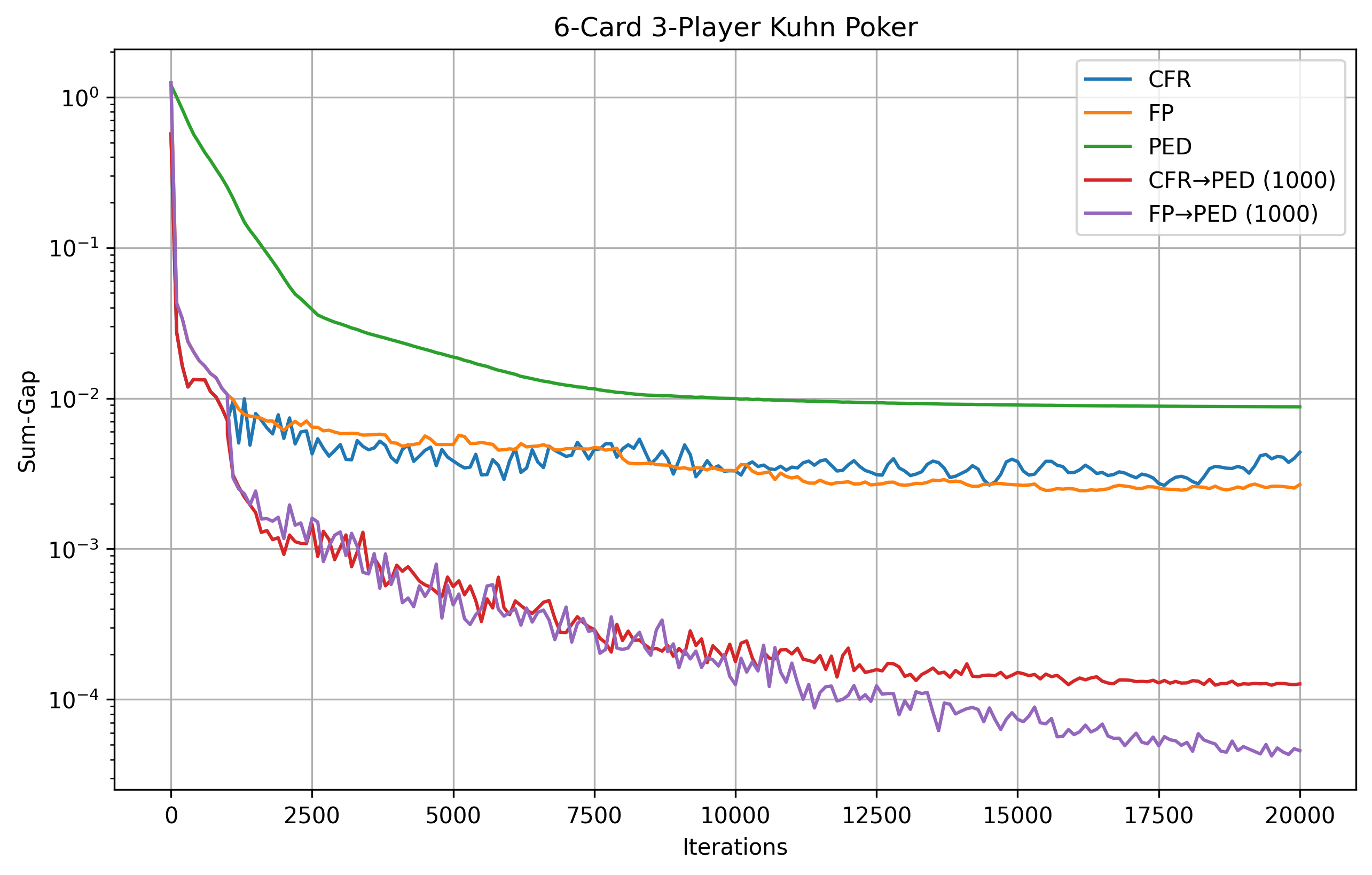}
\caption{Sum gap comparison with best hybrids.}
\label{fig:6c-final-sg}
\end{minipage}
\end{figure*}

We now compare performance of PED to CFR and FP in 3-player 5-card Kuhn poker. Figures~\ref{fig:5c-expl-iter} and \ref{fig:5c-expl-runtime} report the exploitabilities and Figures~\ref{fig:5c-sg-iter}--\ref{fig:5c-sg-runtime} report the $\Phi$ function values, i.e., ``sum gap.'' CFR performed extremely well in the first several hundred iterations before later worsening significantly. FP also reduced exploitability quickly, though with noticeable oscillatory behavior between iterations. In contrast, PED reduced exploitability more gradually during the initial iterations, but unlike FP and CFR its exploitability decreased in a highly stable and nearly monotonic fashion throughout the run. Interestingly, PED also achieved noticeably lower runtime in our implementation, likely because most of its computation consists of vectorized linear-algebra operations in sequence form, whereas FP requires repeated best-response computations and CFR additionally performs recursive game-tree traversals. The results for sum gap were similar, except that PED significantly outperformed FP (while their performances were similar for exploitability), and FP also had noticeably less oscillatory behavior.
                                              
We next repeated experiments on the larger 6-card version, with results in Figures~\ref{fig:6c-expl-iter}--\ref{fig:6c-sg-runtime}. Again, PED decreases both exploitability and sum gap nearly monotonically, though for the 6-card game both FP and CFR consistently outperform PED on both metrics throughout the run. FP outperforms CFR for both metrics for both the 5 and 6-card games; furthermore, FP has lower variability between iterations and a lower runtime. These results indicate that CFR and FP may significantly outperform PED in certain games despite having much larger variability between iterates. They also indicate that CFR and FP perform particularly well in the first several hundred iterations. 

The results suggest that a hybrid approach using FP or CFR during an initial ``burn-in'' phase followed by PED for stable long-run refinement may be particularly effective. Figures~\ref{fig:6c-hybrid-expl} and \ref{fig:6c-hybrid-sg} present results using burn-in phases of 100, 500, and 1000 iterations. Note that we initialize PED with the strategies that produced the smallest exploitability over all burn-in iterations (not necessarily the final iterate), since our objective is to obtain the strongest possible initialization for the refinement phase.
We can see that starting with FP provides consistent improvement over starting with CFR for both metrics, and that using 100 burn-in iterations does the worst for both hybrids. We also observe that using burn-in phases of 1000 does slightly better than 500. For our final comparison, we compare the best CFR-PED and FP-PED hybrids (1000 burn-in iterations) to the original approaches in Figures~\ref{fig:6c-final-expl}--\ref{fig:6c-final-sg}. We can see that FP-PED is the clearly superior approach, achieving exploitability and sum-gap significantly below $10^{-4}.$ Thus, best performance is obtained by combining the strong initial performance of FP with the consistent near-monotonic additional refinement of PED. 
We can alternatively view FP-PED as a two-stage algorithm that uses FP to obtain a strong initialization before applying PED for stable long-run refinement.

\section{Conclusion}
\label{se:conc}
We presented projected exploitability descent (PED), a projected subgradient algorithm for approximating Nash equilibrium in multiplayer imperfect-information games. Rather than optimizing exploitability directly, PED minimizes the smoother surrogate objective $\Phi$ over the sequence-form realization-plan polytope. In our experiments, PED exhibits consistent near-monotonic improvement throughout all runs, while FP and CFR achieve substantially stronger performance during the initial iterations. The hybrid algorithm FP-PED combines the strong initial performance of FP with the stable long-run refinement of PED, achieving the best overall performance in both exploitability and sum gap. Our experiments focused on the generalized $d$-card version of three-player Kuhn poker, a standard benchmark for multiplayer imperfect-information games. In particular, we considered the $d=5$ and $d=6$ variants, which are beyond the reach of existing exact algorithms. 

Future work can evaluate PED on additional game families. Unlike normal-form games, for which there is an established benchmark suite~\cite{Nudelman04:Run}, no comprehensive benchmark suite exists for multiplayer imperfect-information extensive-form games. Prior experiments on multiplayer normal-form games with randomly generated payoffs showed that the NashD projected subgradient algorithm significantly outperformed FP and CFR~\cite{Wang25:Approximating}, and we would expect similar results to hold for analogous experiments on random extensive-form games. A few works have studied a poker variant called 3-player Leduc Hold'em, a game significantly larger than 3-player (4-card) Kuhn poker~\cite{Abou10:Using,MacQueen23:Guarantees}. This game is too large for the sequence-form representation to be stored explicitly in memory, so algorithms that rely on explicit sequence-form representations cannot be applied directly. Future work can explore extensions of our approach to substantially larger games. The sequence-form representation also implicitly assumes that the game has perfect recall, so future work can also explore extensions of PED to games with imperfect recall.

Our best-performing algorithm FP-PED can be viewed as a two-stage algorithm that first uses FP to obtain a strong initialization before warm-starting PED. Prior work has shown that fictitious play itself can achieve much stronger performance for multiplayer Nash equilibrium approximation when run several times with different initializations~\cite{Ganzfried22:Fictitious}. Thus, an interesting direction for future work is to combine multiple randomized initializations of FP with PED by warm-starting PED from the strongest FP trajectory.

\bibliographystyle{plain}
\bibliography{C://FromBackup/Research/refs/dairefs}

\end{document}